\documentclass[pra,twoside,superscriptaddress,twocolumn,floatfix]{revtex4}
\usepackage{times,epsf,amsmath,amssymb,graphicx}
\begin{document}
\title{Multiphoton communication in lossy channels with photon-number
entangled states}
\author{Vladyslav C. Usenko}
\email{usenko@iop.kiev.ua}
\affiliation{Bogolyubov Institute for Theoretical Physics of the National Academy of Sciences, Kyiv, Ukraine}
\affiliation{Dipartimento di Fisica dell'Universit\`a di Milano, Italia.}
\author{Matteo G. A. Paris}
\email{matteo.paris@fisica.unimi.it}
\affiliation{Dipartimento di Fisica dell'Universit\`a di Milano, Italia.}
\date{\today}
\begin{abstract}
We address binary and quaternary communication channels based on
correlated multiphoton two-mode states of radiation in the presence of
losses. The protocol are based on photon number correlations and
realized upon choosing a shared set of thresholds to convert the outcome
of a joint photon number measurement into a symbol from a discrete
alphabet.  In particular, we focus on channels build using feasible
photon-number entangled states (PNES) as two-mode coherently-correlated
(TMC) or twin-beam (TWB) states and compare their performances with that
of channels built using feasible classically correlated (separable)
states. We found that PNES provide larger channel capacity in the
presence of loss, and that TWB-based channels may transmit a larger
amount of information than TMC-based ones at fixed energy and overall
loss. Optimized bit discrimination thresholds, as well as the
corresponding maximized mutual information, are explicitly evaluated as
a function of the beam intensity and the loss parameter.  The
propagation of TMC and TWB in lossy channels is analyzed and the joint
photon number distribution is evaluated, showing that the beam
statistics, either sub-Poissonian for TMC or super-Poissonian for TWB,
is not altered by losses.  Although entanglement is not strictly needed
to establish the channels, which are based on photon-number correlations
owned also by separable mixed states, purity of the support state is
relevant to increase security. The joint requirement of correlation and
purity individuates PNES as a suitable choice to build effective
channels. The effects of losses on channel security are briefly
discussed.
\end{abstract}
\maketitle
\section{Introduction}\label{s:intro}
Communication protocols based on quantum signals have attracted increasing
interest in the recent years, since they offer the possibility of enhancing
either the communication capacity or the security by exploiting the very
quantum nature of the information carriers.  Information may be indeed
conveyed from a sender to a receiver through quantum channels. In order to
achieve this goal a transmitter prepares a quantum state drawn from a
collection of known states and sends it through a given quantum channel.
The receiver retrieves the information by measuring the channel in order to
discriminate among the set of possible preparations and, in turn, to
determine the transmitted signal.  The encoding states are generally not
orthogonal and also when orthogonal signals are transmitted, they usually
lose orthogonality because of noisy propagation along the communication
channel.  Therefore, in general, no measurement allows the receiver to
distinguish perfectly between the signals \cite{Helart, Hel} and the need
of optimizing the detection strategy unavoidably arises.  
\par
A different approach, which will be used in this paper, is to encode
information in the degrees of freedom of a correlated state shared by the
two parties. In this framework, the two parties jointly (and independently)
measure the state and extract the transmitted symbol according to a
previously agreed inference rule.  This kind of schemes, which may be
symmetric or asymmetric depending on the nature of the channels, may serve
either to send a message or to share a cryptographic key.  In particular,
entanglement-based protocols with nonlocal correlations between spatially
separated locations have been proved very effective to provide a pair of
legitimate users with methods to share a secure cryptographic key via
quantum key distribution (QKD).  Besides, the nonclassicality of entangled
states can be used to improve the monitoring of a state against disturbance
and/or decoherence, which, in turn, made entanglement useful to detect
unwanted measurement attempts, {\em i.e} increasing the security of
communication.  Indeed, several quantum-based  cryptographic protocols
\cite{gisin} have been suggested and implemented either for qubits or
continuous variable (CV) systems.
\par
Communication protocols and QKD schemes have been firstly developed for
single qubit \cite{bb84} or entangled qubit pairs \cite{e91}, and
practically implemented using faint laser pulses or photon pairs from
spontaneous parametric downconversion (SPDC) in a pumped nonlinear crystal
\cite{mw}.  Recently, much attention has been devoted to investigating the
use of CV systems in quantum information processing.
In fact, continuous-spectrum quantum variables may be easily manipulated in
order to perform various quantum information processes. This is  the case
of multiphoton Gaussian state of light, {\em e.g.} squeezed-coherent beams
and twin-beams, by means of linear optical circuits, photodetection and
homodyne detection.  In addition, non Gaussian CV states of two modes may
be generated  either by conditional measurements \cite{ng1,ng2} or concurrent
nonlinearities. In turn, CV multiphoton states \cite{cv}, may be used to
increase the effectiveness and reliability of quantum communications and
QKD. Several CV QKD protocols have already been developed on the basis of
quadrature modulations coding of single squeezed \cite{hillery}, coherent
\cite {grosshans2002} and entangled beam pairs
\cite{pereira,ralph,reid,silberhorn}.  Protocols using the sub-shot-noise
fluctuations of photon-number difference of two correlated beams
\cite{funk}, the sub-shot-noise modulations \cite{twb} and the
sub-shot-noise fluctuations of the photon numbers in each of the correlated
modes \cite{tmc2} have been proposed.  Although for CV protocols
unconditional security proofs have not been obtained yet \cite{gott}, they are of
interest and deserve investigations mostly due to the potential gain in
communication effectiveness.
\par
In this paper we address binary and quaternary communication channels
based on photon-number continuous-variable multiphoton entangled states
(PNES), in particular we consider two-mode coherently-correlated (TMC)
or twin-beam  (TWB) states. The communication protocol is based on
photon number correlations and realized upon choosing a shared (set of)
threshold(s) to convert the outcome of a joint photon number measurement
into a symbol from a discrete alphabet.
\par
Notice that, in principle, entanglement itself is not needed to
establish this class of communication channels, which are based on
photon-number correlations owned also by separable mixed states.  On the
other hand, purity of the support state is relevant to increase security
of the channel. In fact, if the information is encoded in classically
correlated (mixed) states an intruder can easily measure the number of
photons in either of the modes and then recreate the mixed state mode by
activating the corresponding number of single-photon or photon-pair
sources (the latter case with the degenerate SPDC process is already
quite realistic).  Thus the information encoded in the photon number of
a mixed state mode can be effectively intercepted. On the other hand,
this attack is not effective in case of PNES-based channels, since the
destruction of the mutual second order coherence of a PNES state can be
revealed by a joint measurement on the two modes, which can be
accomplished if the receiver, instead or besides extracting the bit
value, randomly sends his mode or part of it back to the sender to let
her check the presence of an eavesdropper, analogously to "two-way"
quantum cryptography based on individually coherent entangled beams
\cite{twoway}.   For the PNES-based protocols although no strict proofs
of security can be offered, TMC-based protocols may be proved secure
against realistic intercept-resend eavesdropping. The security mostly
relies on the fact that the generation of traveling Fock states of
radiation, despite several theoretical proposals based on tailored
nonlinear interactions \cite{non}, conditional measurements \cite{con},
or state engineering \cite{eng,fockfil}, is still extremely challenging
from the experimental point of view.  Overall, it is the joint
requirement of correlation and purity that leads to individuate PNES as
a suitable choice for building effective and, to some extent, secure
communication channels.  
\par
The main goal of the paper is twofold. On the one hand we consider
communication channels based on realistic class of PNES and analyze the
effects of losses on the performances of the protocol. On the other
hand, we optimize the performances of our protocol and compare the
results of PNES-based schemes to that obtained using a realistic kind of
classically correlated (mixed, separable) states as a support. The
evolution of TMC and TWB in lossy channels, as well as that of
classically correlated states, is analyzed to calculate the joint photon
number distribution and evaluate the survival of correlations.  Using
this results we determine the optimized bit discrimination thresholds
and the corresponding channel capacity (maximized mutual information)
for binary and quaternary alphabets. The effects of losses on security
of the protocols against intercept-resend attacks are briefly discussed.  
\par
The paper is structured as follows: in Section \ref{s:pnes} we
describe the communication protocol and introduce the correlated
states, either PNES or classically correlated, that will be
considered as a support. In Section \ref{s:loss} we analyze the
propagation of the above states in a lossy channel, and evaluate
the joint photon number distribution and the correlations. In
Section \ref{s:IM} we optimize the bit discrimination threshold
for binary and quaternary alphabets and evaluate the
corresponding channel capacity.  Finally, in Section \ref{s:outro} 
we briefly discuss security and close the paper with 
some concluding remarks.
\section{Communication channels based on photon-number correlations}
\label{s:pnes}
Quantum optical communication channels can be established using
multiphoton entangled states of two field-mode, which provide the 
necessary correlations between  the two parties. In this work we 
investigate the information capacity of quantum channels built using 
as a support a specific class of bipartite entangled states, which 
we refer to as photon-number entangled states (PNES). In the Fock 
number basis, PNES may be written as 
\begin{equation}
\label{eq:twin}
\left| \Psi \right\rangle\rangle =
\sum\limits_n {c_n \left| {n,n} \right\rangle\rangle } ,
\end{equation}
where $\left| {n,n} \right\rangle\rangle = \left| n \right\rangle_1 
\otimes \left| n  \right\rangle _2$ and $\sum_ |c_n |^2 =1$. 
As we will see an effective channel may be established exploiting the 
strong correlation between the photon number distributions of the
two modes. Indeed, PNES show perfect (total) correlations in the photon 
number, {\em i.e.} the correlation index
\begin{align}\label{gamma}
\gamma = \frac{\langle n_1 n_2\rangle - \langle n_1 \rangle 
\langle n_2 \rangle}{\sqrt{\sigma_1^2 \sigma_2^2}}\:,
\end{align}
with $n_j = a^\dag_j a_j$, $j=1,2$ and $\sigma_j^2=\Delta n_j^2$, is equal 
to one for any PNES. On the other hand, the degree of entanglement
strongly depends on the profile $c_n$.
PNES may be generated by means of parametric optical oscillator (OPO) 
exploiting seeded PDC process in a nonlinear crystal placed in and optical
cavity \cite{walls}. Several implementations have already been reported, with 
the generation of PNES with photon number statistics varying from
super-Poissonian to sub-Poissonian after post-selection \cite{twin1, twin2, twin3}.
Meanwhile, several quantum communication schemes and QKD protocols were
proposed using PNES, with information encoded in the beam intensity
\cite{twb, tmc} or intensity difference \cite{funk,zhang}. 
\par
The bits coding/decoding for a PNES-based communication protocol is rather
natural: in the binary case each of the legitimate users measure the incoming 
photon number in a predetermined time slot and compare the obtained value to 
a given bit threshold. If the detected value is above the threshold the 
corresponding bit value is assigned to one,  zero otherwise
\begin{equation}
\label{eq:protocol}
B_2 = \left\{ 
\begin{array}{*{20}c}
n \le T \to 0 \hfill \\
n > T   \to 1 \hfill \\
\end{array}  
\right.\:.
\end{equation}
The scheme may be also extended to a $M$-letter protocol by introducing different 
thresholds
\begin{equation}
\label{eq:protocolK}
B_M = \left\{ 
\begin{array}{ccc}
n \le T_1 & & \to 0 \hfill \\
T_1 < n \le T_2 & &   \to 1 \hfill \\
T_2 < n \le T_3 & &   \to 2 \hfill \\
... & & ...\hfill \\ 
n > T_{M-1} & & \to M 
\end{array}  
\right.\:.
\end{equation}
The effectiveness of these generalizations depends of course on the beam
intensity and on the resolution thresholds of the detectors. 
PNES-based protocols of this kind have been suggested
\cite{tmc2} and analyzed in the ideal case using two-mode
coherently correlated (TMC) states  using, in the the binary case, a threshold value 
equal to the integer part of the average photon number.
In the Fock basis TMC states \cite{agarwal1,agarwal2} are written as follows 
\begin{equation}
\label{eq:TMC}
\left| \lambda \right\rangle\rangle = \frac{1}{\sqrt {I_0 \left( {2\left| \lambda 
\right|} \right)} }\sum\limits_n {\frac{\lambda ^n}{n!}\left| {n,n} 
\right\rangle\rangle } 
\end{equation}
where $\lambda \in {\mathbb C}$ and $I_0(x)$ denotes a modified Bessel' 
function of the first kind. Without loss of generality we will consider
$\lambda$ as real throughout the paper.  
The average photon number of the state 
$|\lambda\rangle\rangle$ is given by 
$$N_\lambda=
\frac{2 \lambda I_1 (2\lambda)}{I_0(2\lambda)}\:.
$$
TMC are eigenstates of the product of the annihilation operators of the 
two radiation modes $a_1 a_2 |\lambda\rangle\rangle = \lambda
|\lambda\rangle\rangle$ and for this reason they are also referred to as
pair-coherent states. The two partial traces 
\begin{align}
\rho_1 & \equiv \hbox{Tr}_2 [|\lambda\rangle\rangle\langle\langle \lambda |] = 
\rho_2 \equiv \hbox{Tr}_1 [|\lambda\rangle\rangle\langle\langle \lambda |] 
\nonumber \\ &= \frac{1}{I_0 (2 \lambda )} \sum\limits_n \: \frac{\lambda ^{2n}}{n!^2}\:
|n\rangle\langle n | \label{TMCpt}\;,
\end{align}
show sub-Poissonian photon statistics. In fact, the Mandel parameter
$Q = \sigma^2/ \left\langle n \right\rangle - 1$
is given by 
\begin{align}
Q_\lambda = \lambda \frac{I_0^2(2\lambda)-I_1^2(2\lambda)}{I_0(2\lambda)I_1(2\lambda)} -1
\end{align}
and it is negative for any value of $\lambda$.
\par
A communication channel based on TMC relies on the strong photon number
correlations, which allow to decode a random bit sequence by carrying
out independent and simultaneous intensity measurements at two remote
locations.  On  the other hand, the security of the scheme is based on
checking the beam statistics coming from the measurement results against
the (sub-Poissonian) expected one. It was shown that any realistic
eavesdropping attempts introduce perturbations that are significant
enough to be detected, thus making eavesdropping ineffective
\cite{tmc2}. In addition, the extension to $M=2^d$-letter  alphabets was
shown to be effective, {\em i.e} increase the information capacity to
$d$ bits per measurements, also making the protocol more secure against
intercept-resend attacks \cite{tmc}.   
\par
Another relevant class of PNES is given by the so-called twin-beam state (TWB)
\begin{equation}
\label{eq:twb}
| x \rangle\rangle = \sqrt {\left( {1 - x^2} \right)} \sum\limits_n 
{x^n\left| {n,n} \right\rangle\rangle }.
\end{equation}
where $x\in{\mathbb C}$ and $0\leq |x|\leq 1$, which are entangled
two-mode Gaussian states of the field and represents the crucial
ingredient for CV teleportation and dense coding.  Without loss of
generality we assume $x$ as real, the average photon number of TWB is
thus given by 
$$ N_x= \frac{2 x^2}{1-x^2}\:.$$
The two partial traces of $|x\rangle\rangle$ are equal to thermal states
$\nu_{\frac N2} = (1+N/2)^{-1} [N/(2+N)]^{a^\dag a}$ with $N/2$ average
photon number each. The Mandel parameter is positive and equal to
$Q=N/2$ for both the modes.  As we will see, also TWB may employed to
built a CV communication protocol analogue to that described above. The
channel capacity is generally larger than for TMC, though the
super-Poissonian statistics of the partial traces make the security
issue  more relevant. 
\par
\begin{figure}[h]
\begin{tabular}{ll}
\includegraphics[width=0.2\textwidth]{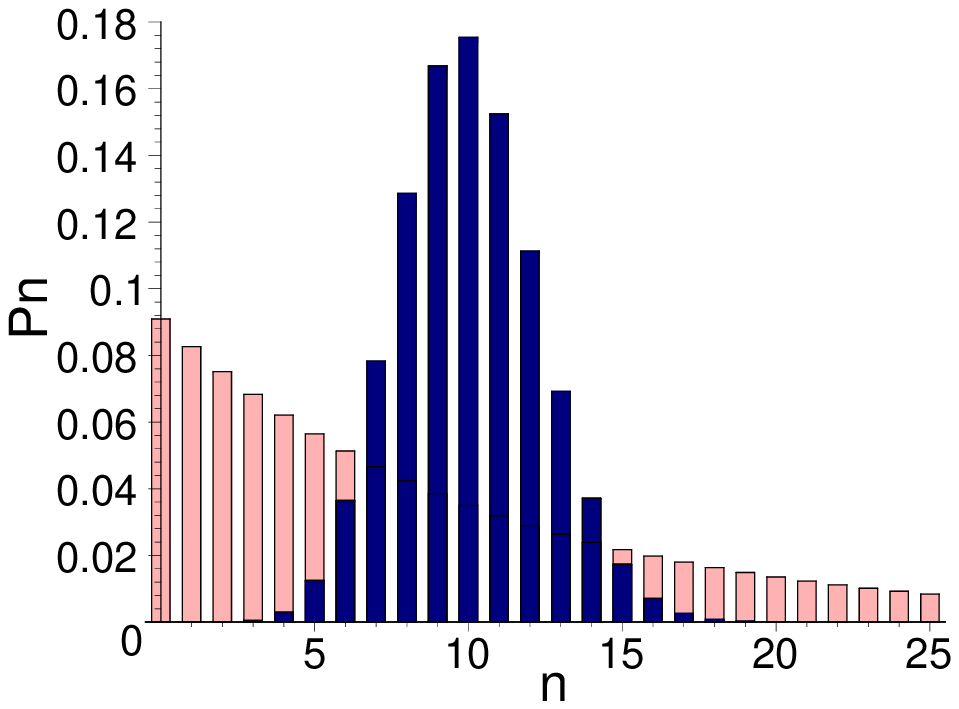}
\includegraphics[width=0.2\textwidth]{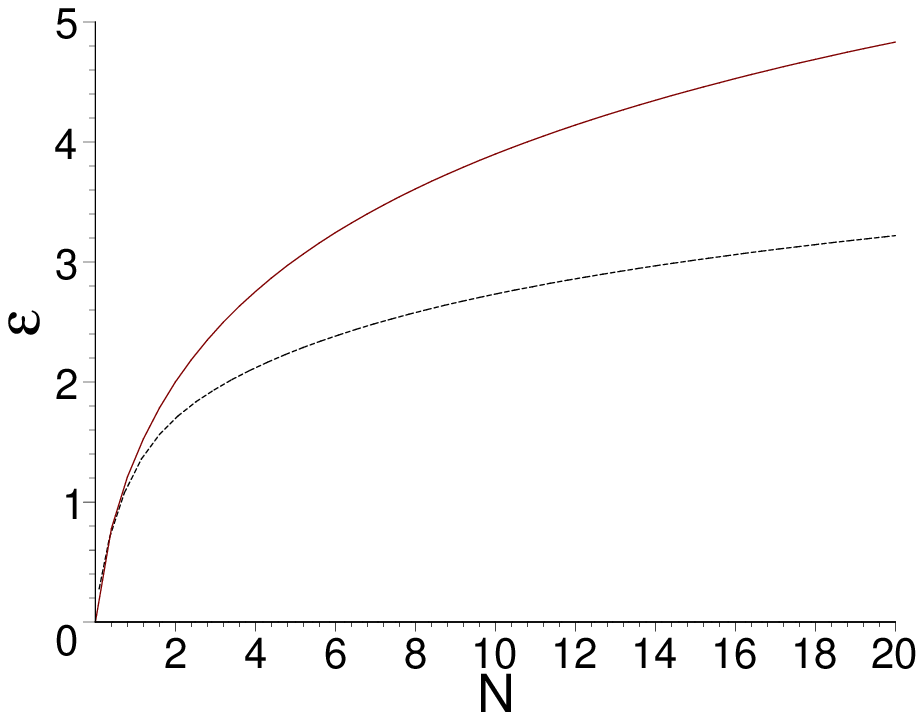}	
\end{tabular}
\caption{(Color online) Left: comparison between the mode photon number 
profiles $P_n \equiv |c_n|^2$ of TMC (dark grey) and TWB (light grey) with state average 
photon number $N=20$; Right: 
entanglement (VN entropy of the partial traces) of TMC (lower, dotted line) 
and TWB (upper, solid line) states as a function of the state average photon number $N$. 
\label{TMCvsTWB}}
\end{figure}

In Fig. \ref{TMCvsTWB} we show the typical photon number distribution
$|c_n|^2$ of TMC' and TWB' partial traces. We also report the degree of
entanglement of the  two states, evaluated as the Von-Neumann entropy
$S[\varrho]=-\hbox{Tr}[\varrho\log\varrho]$ of the partial traces, {\em
i.e.} $\epsilon= S[\varrho_1] =S[\varrho_2]$, as a function of the
average photon number. Notice that TWBs show larger entanglement; indeed
they are maximally entangled states for a CV two-mode system at fixed
energy \cite{visent}.
\subsection{Classically correlated states}
As already mentioned in the introduction, the communication protocols expressed 
by Eqs. (\ref{eq:protocol}) and (\ref{eq:protocolK}) may
be, in principle, implemented also without entanglement, {\em e.g.} using
mixed separable states of the form 
$$R = \sum_{n=0}^{\infty} P_{nm} \left|{n \times n }\right| 
\otimes \left|{m \times m }\right|\:,$$ 
with any non factorized profile $P_{nm}\neq p_n p_m$, as for example
$P_{nm}= \delta_{nm}|c_n|^2$, $c_n$ being the amplitude of a PNES.  This
kind of classically correlated separable mixed states may have the
necessary correlations to establish the channel but, on the other hand,
have serious disadvantages compared to PNES in terms of security of the
quantum channel.  Thus in the following we will consider classically
correlated states to assess the improvement of the channel capacity
using PNES, however keeping in mind that any kind of classically
correlated mixed state does not provide reliable security of the
transmitted information.
\par
As a paradigm of classically correlated state we consider the state
obtained at the output of a balanced beam splitter fed by a thermal
state (and with the second port unexcited). This is a feasible class 
of states, which we refer to as two-mode thermal (TTH) states, that 
have been proved very effective to implement the so-called ghost 
imaging by classical means. In turn, this opened a debate about the 
usefulness of entanglement in protocols based on photon correlations
\cite{gh1,gh2}.  As we will see, although  TTH perform well in ghost 
imaging, PNES-based communication protocols achieve a larger channel 
capacity. 
\par
The density matrix of TTH is given by 
\begin{align}
R_H = U_{\frac{\pi}{4}} \nu_N \otimes |0\rangle\langle 0| U_{\frac{\pi}{4}}^\dag
\label{TTH}\;, 
\end{align}
where $U_{\frac{\pi}{4}}=\exp\{\pi/4 (a^\dag b + b^\dag a)\}$ 
is the evolution operator of the balanced beam splitter and $\nu_N$ 
is a thermal state with $N$ average photons.
$R_H$ is a mixed separable Gaussian state with correlation index 
given by $\gamma=N/(N+2)$. For large $N$ the correlation index
approaches one, in agreement with the strong correlations observed 
in ghost imaging experiments. 
The partial traces of $R_H$ are both thermal states with $N/2$ average 
photons and Mandel parameter $Q=N/2$, whereas the the two-mode photon 
distribution is given by 
\cite{JointDiff}
\begin{align}
P_H(p,q) = \frac{1}{1+N} 
\left(\begin{array}{c} p+q \\ p\end{array} \right)
\left[\frac{N}{2(1+N)}\right]^{p+q} \:. 
\end{align}
\section{Correlated states in a lossy channel}\label{s:loss}
\begin{figure}
\includegraphics[width=0.48\textwidth]{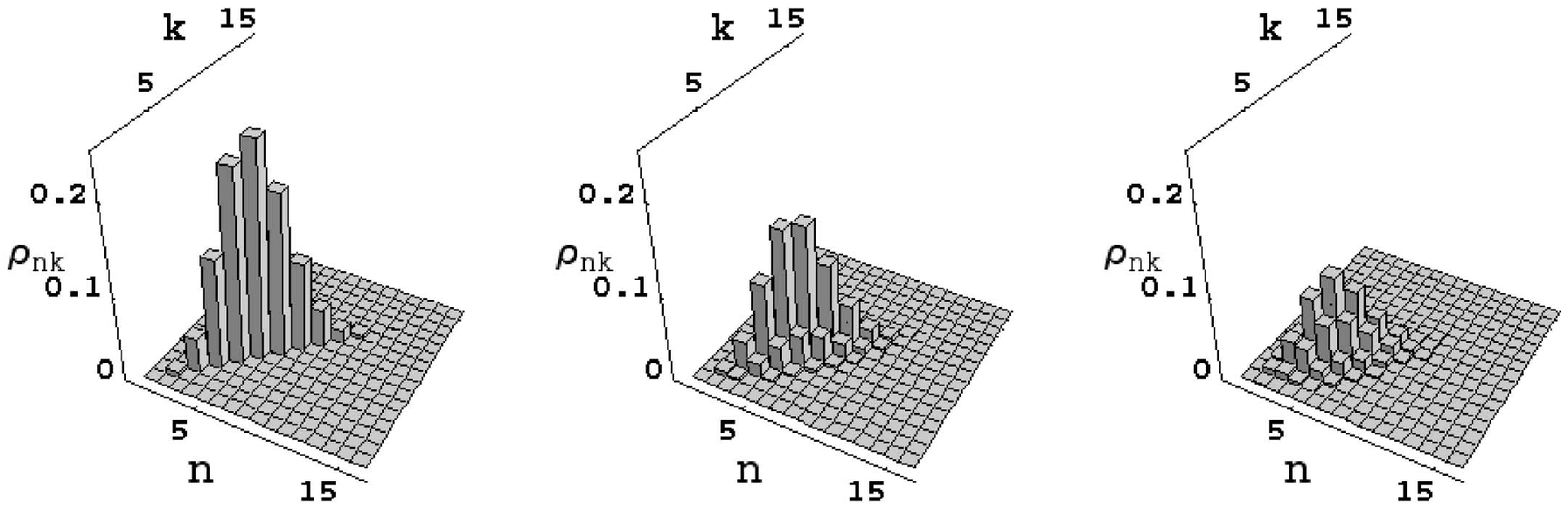}
\includegraphics[width=0.48\textwidth]{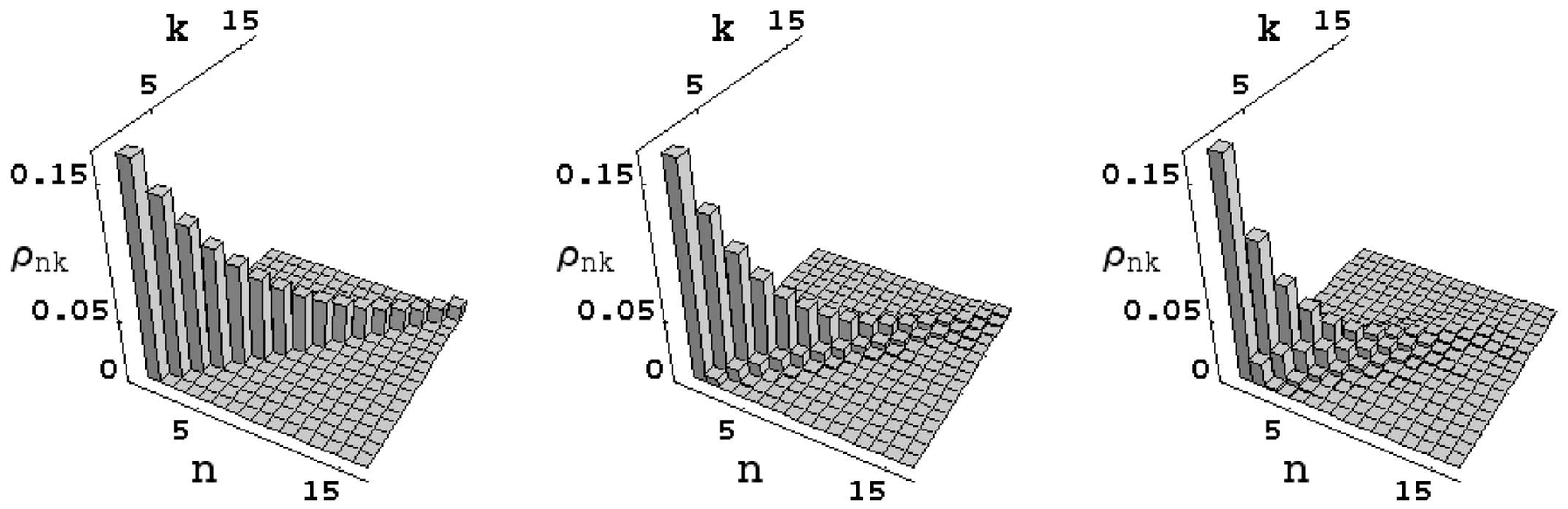}
\caption{Effect of losses on the joint photon-number distribution $P_\eta (p,q)$ 
of TMC and TWB with average photon number $N=10$. The plots on the top line refer to TMC, on the bottom line to TWB. 
From left to right the distributions for $\eta=1$ (no loss), $\eta=0.95$ and
$\eta=0.85$.}\label{f:pdns}
\end{figure}

In order to assess the performances of our protocol in realistic 
conditions we investigate the propagation of the support states 
in lossy optical media, as the evolution in a fiber. We model the 
loss mechanism by the standard quantum optical Master equation, {\em i.e.} 
as the interaction with a bath of oscillators. We also assume that the 
noisy environment is acting independently on the two modes. At zero 
temperature the evolution of a two-mode state described by the density 
matrix $R$ is given by 
\begin{equation}\label{ME}
\dot {R} = \left ( L[a_1] + L[a_2] \right) \: R  
\:,
\end{equation}
where dots denote time derivative and the Lindblad superoperator 
$L[a]$ acts as follows
\begin{equation}
L[a]\varrho= \frac{\Gamma }{2}(2a\rho a^ + - a^ + a\rho - \rho a^ + a)
\:.
\end{equation}
Assuming that $R_0$ denotes the 
initial density matrix, the evolved state, {\em i.e.} the solution 
of the Master equation (\ref{ME}), is given by
\begin{equation}
\label{eq:reta}
R_\eta = \sum\limits_{n,k = 0}^\infty 
A_n^{(1)} A_k^{(2)} R_0 A_k^{(2)\dag} A_n^{(1)\dag},
\end{equation}
where the elements of the maps are given by 
\begin{align}
\label{eq:an}
A_n^{(j)} &= \frac{\left( {\eta_j^{ - 1} - 1} \right)^{n / 2}}{\sqrt {n!} }a_j^n\eta_j 
^{\frac{1}{2}a_j^\dag a_j} \quad j=1,2
\end{align}
and $\eta_j = \exp ( - \Gamma_j t)$ will be referred to as the
loss parameter.
The evolution of TMC and TWB corresponds to the evolution of pure states 
of the form  $R_0 = \left. {\left| {\psi _0 } \right\rangle } \right\rangle \left\langle 
{\left\langle {\psi _0 } \right|} \right.$ with $\left| {\psi _0 } \right\rangle\rangle$ 
being the PNES of Eq. (\ref{eq:twin}). 
The joint photon number distribution after the propagation corresponds 
to the diagonal matrix elements of the evolved state 
$P_\eta (p,q)=\langle \langle p,q |R_\eta |p,q \rangle\rangle$. 
Assuming that the coefficients $c_n$ in (\ref{eq:twin}) are real and using
Eqs. (\ref{eq:reta}) and (\ref{eq:an}) we arrive at 
\begin{align}\label{peta}
P_\eta \left( {p,q} \right) = 
\sum_{n,k} \sum_{i,j} &c_i  c_j \left\langle p \right|A_n^{(1)} 
\left| i \right\rangle \left\langle q \right|A_k^{(2)} \left| i \right\rangle 
\nonumber \\ & \times
\left\langle j \right|A_n^ {(1)\dag} \left| p \right\rangle \left\langle j 
\right|A_k^ {(2)\dag} \left| q \right\rangle 
\end{align}
with
\begin{equation}
\left\langle p \right|A_n^{(j)} \left| i \right\rangle = \frac{\left( {\eta_j ^{ - 
1} - 1} \right)^{n / 2}}{\sqrt {n!} }\eta_j ^{\frac{p + n}{2}}\sqrt 
{\frac{\left( {p + n} \right)!}{p!}} \delta _{i,p + n} 
\end{equation}
and analogously for the other terms.
Upon substituting in Eq. (\ref{peta}) the expression
of the coefficients $c_n$ for the TMC and TWB we obtain
the output joint photon distributions 
\begin{widetext}
\begin{align}
\label{petaTMC}
P_{\lambda,\eta_1,\eta_2} \left( {p,q} \right) & = 
\left(
I_0 \left({2\left| \lambda \right|} \right)\: p!\,q!
\right)^{-1}
I_{|p-q|} \left[
{2\left| \lambda \right|\sqrt{(1 - \eta_1)(1 - \eta_2)} } \right]
\lambda ^{p +
q}\eta_1^p\eta_2^q(1-\eta_1)^{\frac{q-p}{2}}(1-\eta_2)^{\frac{p-q}{2}}
\\
\label{petaTWB}
P_{x,\eta_1,\eta_2} \left( {p,q} \right)  & =  \left( {1 - x^2}
\right)\left({\frac{\eta_1}{1-\eta_1}}\right)^p\left({\frac{\eta_2}{1-\eta_2}}\right)^q
\left[{x^2\left({1-\eta_1}\right)\left({1-\eta_2}\right)}\right]^M \left(
{{\begin{array}{*{20}c}
 M \hfill \\
 m \hfill \\
\end{array} }} \right) \cdot {} 
\nonumber\\
 & \times { }_2F_1 \left[ {\left\{ {1 + M,1 + M} \right\},\left\{ {1 + d}
\right\};x^2\left( {1 - \eta_1 } \right)\left( {1 - \eta_2 } \right)} \right],
\end{align}
\end{widetext}
where $d = \left| {p - q} \right|$, $M = \max \left( {p,q} \right)$, $m = \min
\left( {p,q} \right)$, $I_d(x)$ is the $d$-th modified Bessel
function of the first kind, and ${ }_2F_1 \left[ {\left\{ {a,b} \right\},\left\{ c
\right\};x} \right]$ denotes a hypergeometric function.
Eq. (\ref{peta}) can be easily generalized to mixed input: for TTH states 
the evolution of the state $R_H$ corresponds to a
joint photon number distribution given by 
\begin{align}\label{petaTTH}
P_{N,\eta_1,\eta_2}(p,q) = 2\eta_1^p\eta_2^q
\:\frac{N^{p+q}}{
\left[{2+N\left({\eta_1+\eta_2}\right)}\right]^{p+q+1}} 
\:\binom{p+q}{p} \end{align}
The correlation index $\gamma$ decreases with losses. Upon the 
evaluation  of the first moments using Eqs. (\ref{petaTMC}), (\ref{petaTWB})
and (\ref{petaTTH}) we arrive at 
\begin{align}
   \gamma_\lambda & = \sqrt{\eta_1 \eta_2}
\label{gml}
\\ \gamma_x & = \frac{(2+N_x) \sqrt{\eta_1 \eta_2}}{\sqrt{(2 + N_x
\eta_1)(2+N_x \eta_2)}}
\label{gmx}
\\ \gamma_\nu & = \frac{N \sqrt{\eta_1 \eta_2}}{\sqrt{(2 + N
\eta_1)(2+N \eta_2)}}
\label{gmt}
\end{align}
For TMC the correlation index does not depends on the
input energy.
In Fig. \ref{f:pdns} we show the joint photon-number distribution
of TMC and TWB for different loss parameter and $N=10$.
We also notice that the Mandel parameter of the partial traces shows a simple rescaling 
$Q_j \rightarrow \eta_j Q_j$ and thus the sub-Poissonian statistics of TMC, 
and the super-Poissonian one of TWB are not altered by the propagation.
\section{Optimized bit thresholds and channel capacities}
\label{s:IM}
\subsection{Symmetric channels}
Once the joint probability distribution is known we may evaluate
the mutual information between the two parties and optimize it 
against the threshold(s) for the different channels. Let us illustrate 
the procedure for the case of a binary alphabets assuming a symmetric lossy 
channel {\em i.e.} $\eta_1=\eta_2=\eta$. The effects of asymmetric will be 
discussed in the next sub-Section. Upon adopting the decoding rule (\ref{eq:protocol}) 
the two parties infer the same symbol with probabilities
\begin{align}
p_{00} &= \sum_{p=0}^T \sum_{q=0}^T P_\eta (p,q) \\
p_{11} &= \sum_{p=T}^\infty \sum_{q=T}^\infty P_\eta (p,q)
\label{p00p11}\;.
\end{align}
In the ideal case, {\em i.e.} with no losses, PNES-based 
protocols achieve $p_{00}+p_{11}=1$, due to perfect correlations
between the two modes. On the other hand, if $\eta\neq 1$ the unwanted 
inference events "01" and "10" may occur with probabilities
\begin{align}
p_{01} &= \sum_{p=0}^T \sum_{q=T}^\infty P_\eta (p,q) \\
p_{10} &= \sum_{p=T}^\infty \sum_{q=0}^T P_\eta (p,q)
\label{p10p01}\;.
\end{align}
The probabilities are not independent since the normalization 
condition $p_{00}+p_{10}+p_{01}+p_{11}=1$ holds.  
The mutual information between the two alphabets reads as follows
\begin{align}
I_2 = \sum_{i=0}^1 \sum_{j=0}^1 p_{ij} \log \frac{p_{ij}}{q_i r_j}
\label{IM}\;,
\end{align}
where 
\begin{align}
q_i=&p_{i0}+p_{i1} \quad i=0,1 \\ 
r_j=&p_{0j}+p_{1j} \quad j=0,1 \:,
\end{align}
represents the marginal probabilities, {\em i.e.} the unconditional 
probabilities of inferring the symbol ``i'' (``j'') for the first 
(second) party. The mutual information, once the average number of 
input photons and the loss parameter have been set, depends only on 
the threshold value $T$. The channel capacity ${\cal C}_2=\max_T I_2$ 
corresponds to the maximum of the mutual information over 
the threshold.
\begin{figure}[h]
\begin{tabular}{lll}
\includegraphics[width=0.15\textwidth]{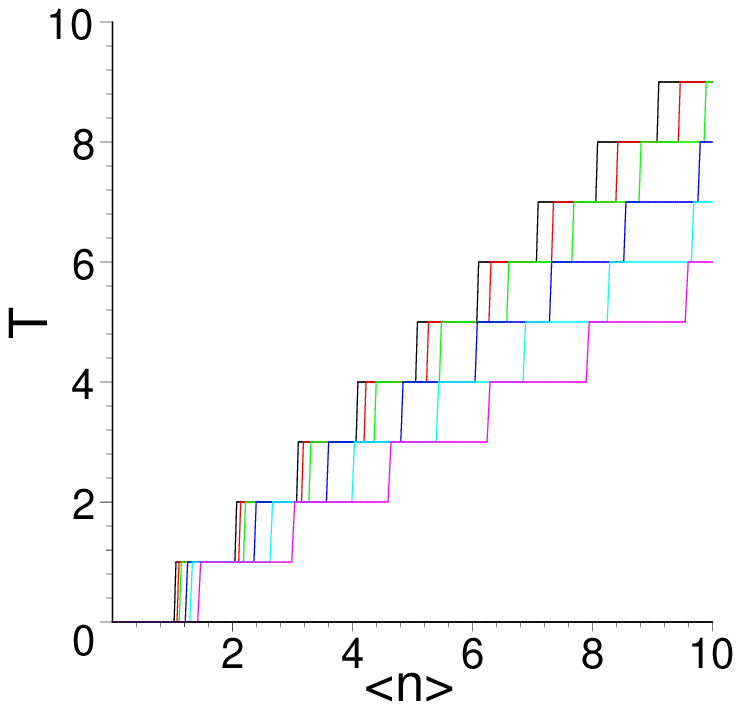}&
\includegraphics[width=0.15\textwidth]{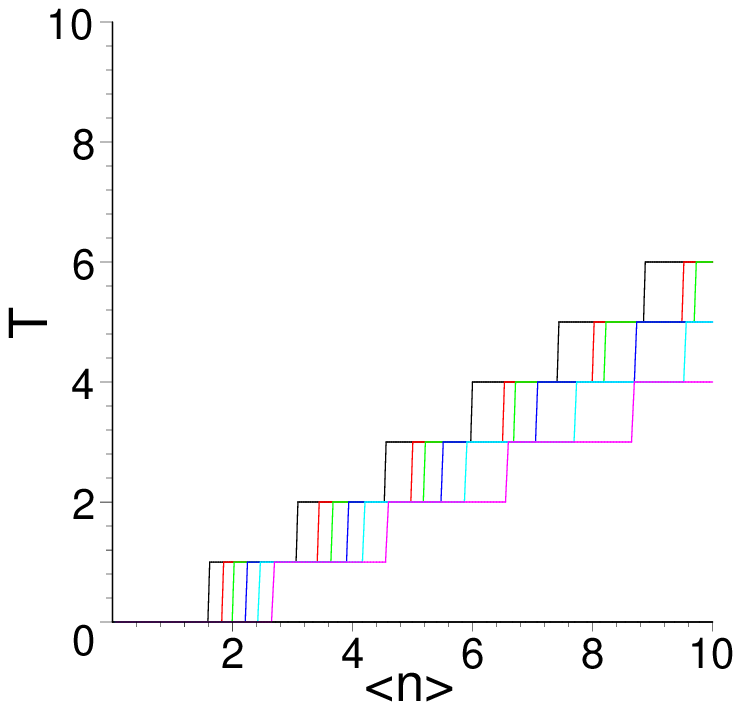}&
\includegraphics[width=0.15\textwidth]{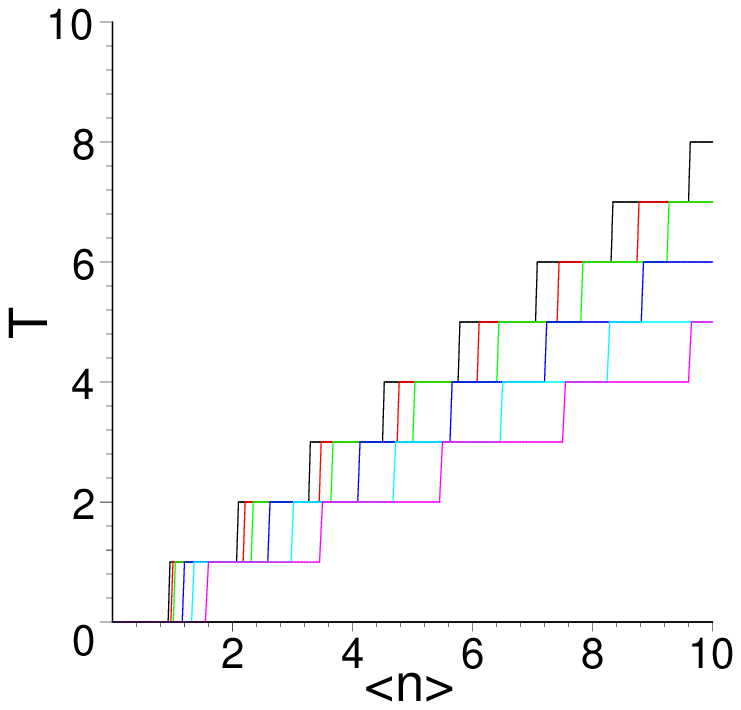}\\ 
\includegraphics[width=0.15\textwidth]{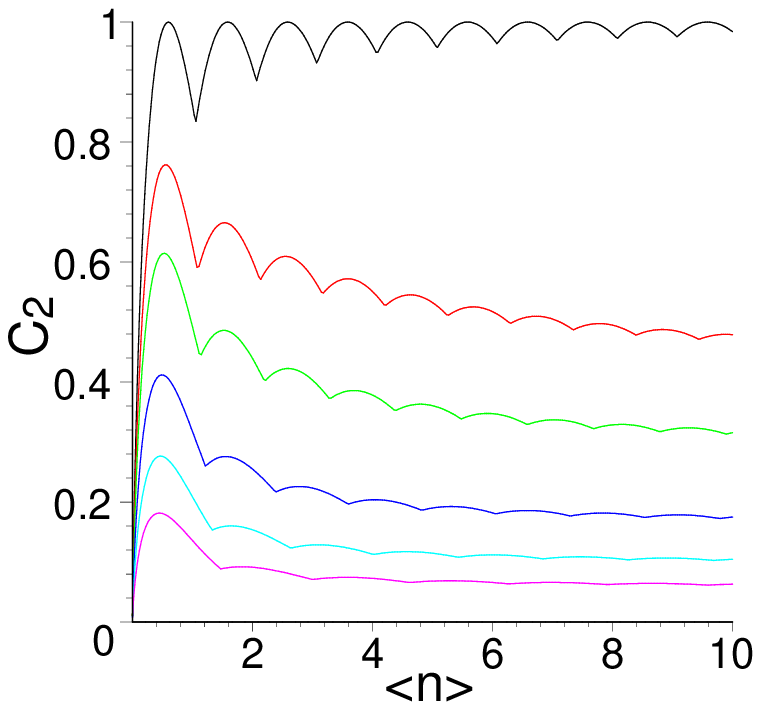} &
\includegraphics[width=0.15\textwidth]{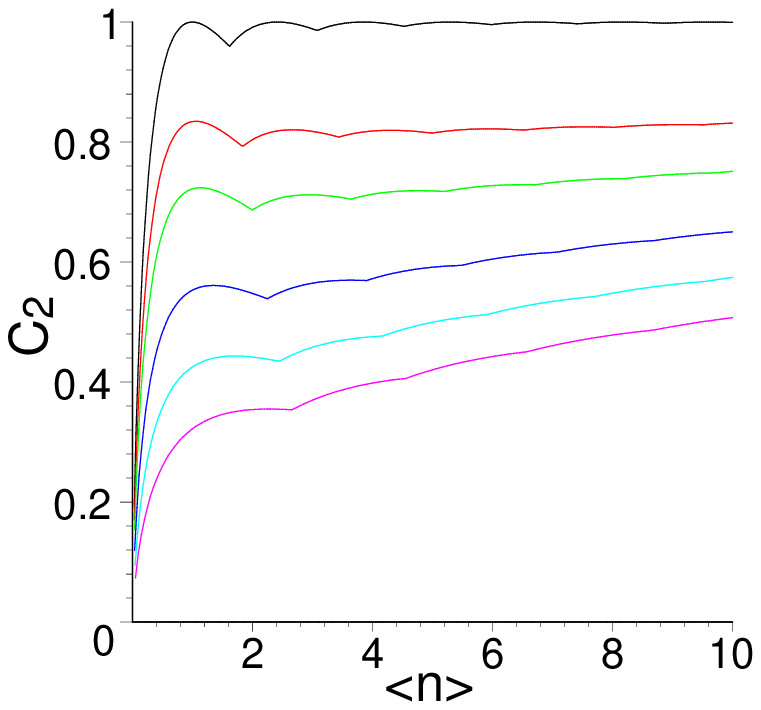} & 
\includegraphics[width=0.15\textwidth]{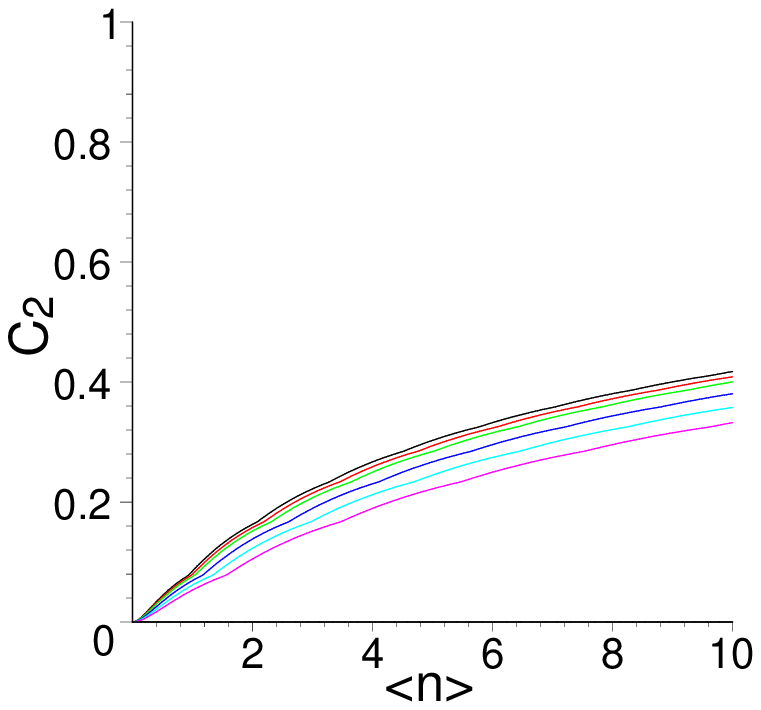}
\end{tabular}
\caption{(Color online) Optimized threshold (top) and channel capacity (bottom) 
for TMC-, TWB- and TTH-based 2-letter protocols as a function of the 
mode average 
photon number at the input and for different values of the loss parameter
$\eta$ (symmetric channels). In all the plots, from bottom to top: 
$\eta=0.6, \eta=0.7, \eta=0.8, 
\eta=0.9, \eta=0.95$ and $\eta=1$, the upper curves correspond to the ideal
case with no losses. \label{2-letter}}
\end{figure}
\par
The mutual information has been maximized numerically by looking for the
optimal bit discrimination threshold as a function of the input energy.
The optimal thresholds and the corresponding channel capacities are
shown in Fig. \ref{2-letter}. Notice that the threshold for TMC only
slightly increases with loss and it is always larger than the TWB's one.
On the other hand,  the threshold for TTH is smaller, and the resulting
channel capacity is smaller than for PNES-based schemes as far as the
loss is not too strong.  At fixed energy the channel capacity is larger
for TWB than for TMC.  Notice that the capacity for a single-mode
two-letter intensity modulation/direct detection (IMDD) channel is always
smaller than ${\cal C}_2$ in the (relevant) low photon-number regime
\cite{coment}. IMDD reaches the unity effectiveness for the mode
average photon number $\langle n \rangle \approx 6$, giving just about $0.04$ bits for 
$\langle n \rangle \approx 1$, while
PNES schemes perform maximally already at $\langle n \rangle \approx 1$. 
Thus the PNES-based protocols are more convenient in
terms of energy expenditure.
\par
\begin{figure}[h!]
\begin{tabular}{lll}
\includegraphics[width=0.15\textwidth]{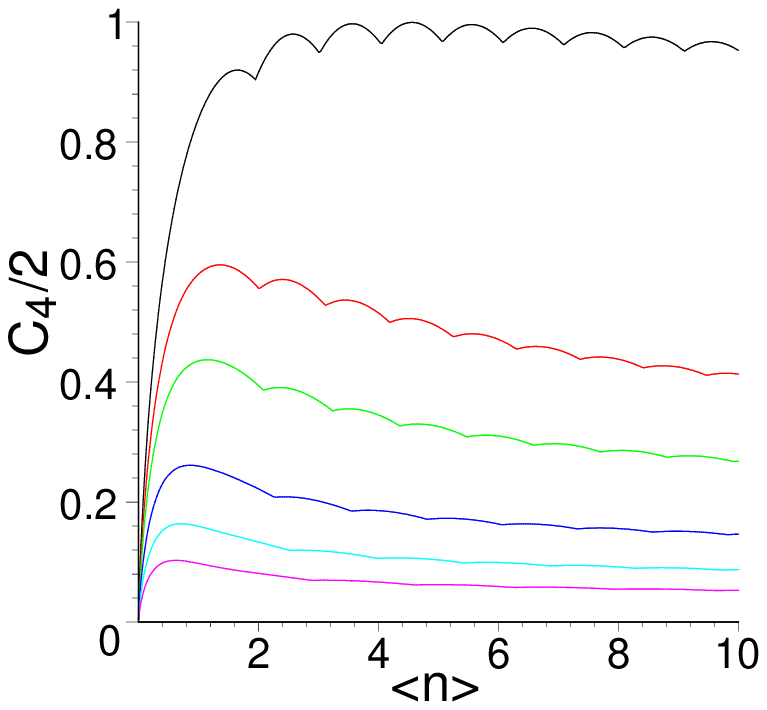} &
\includegraphics[width=0.15\textwidth]{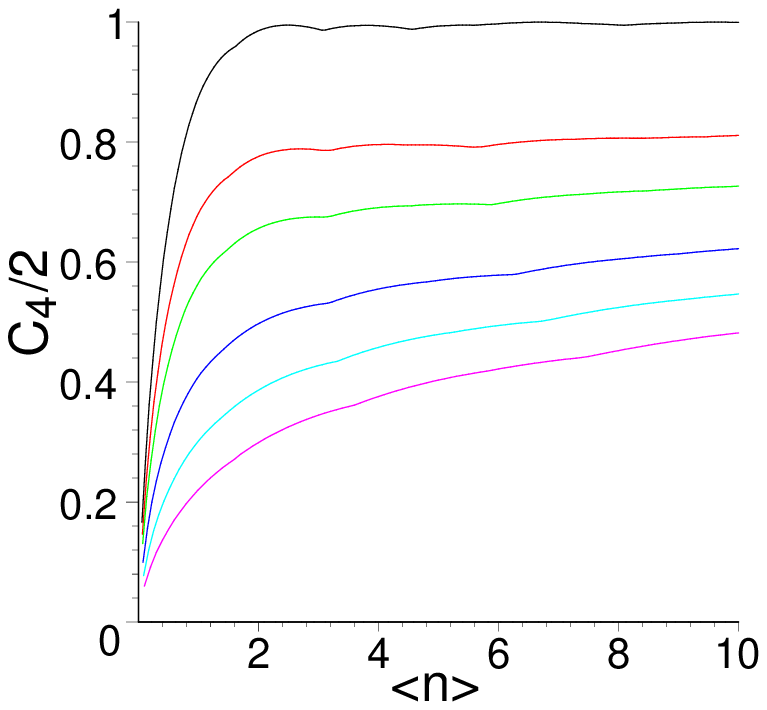} & 
\includegraphics[width=0.15\textwidth]{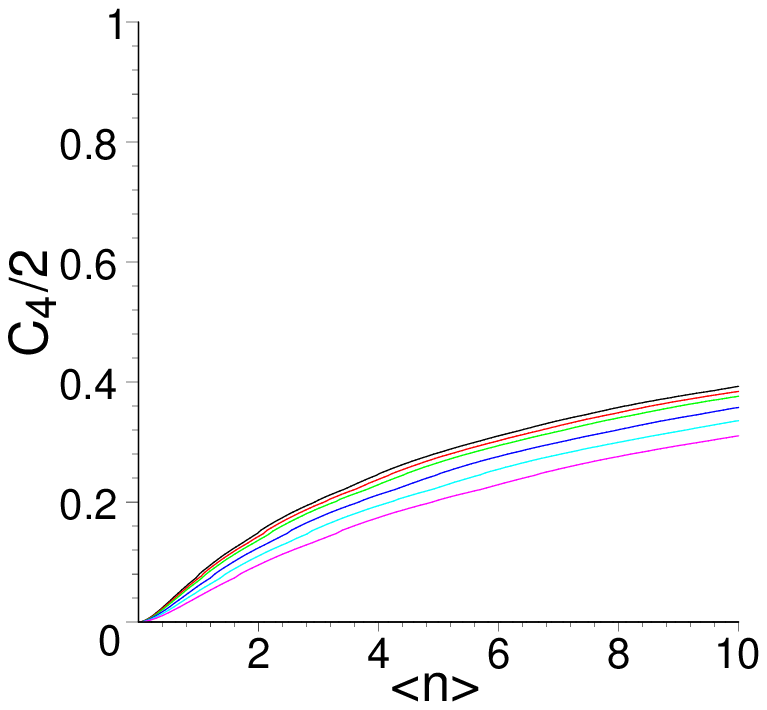}
\end{tabular}
\caption{(Color online) Optimized normalized channel capacity  
for TMC-, TWB- and TTH-based 4-letter protocols as a function of the mode average 
photon number at the input and for different values of the loss parameter
$\eta$ (symmetric channels). In all the plots, from bottom to top: 
$\eta=0.6, \eta=0.7, \eta=0.8, 
\eta=0.9, \eta=0.95$ and $\eta=1$, the upper curves correspond to the ideal
case with no losses. \label{4-letter}}
\end{figure}

The channel capacity for a PNES based $M$-letter protocol may be
analogously derived by maximizing the mutual information versus the
$M-1$ bit thresholds.  In the ideal case (no loss) the capacity
obviously increases with $M$ (as $\log_2 M$). A question arises on
whether this effect is robust against losses.  In Fig. \ref{4-letter} we
report the rescaled capacity ${\cal C}_4/2$ as it results from the
threshold optimization of the $4$-letter TMC-, TWB- and TTH-based
protocols respectively. Besides the dubbing of the capacity due to the
alphabet dimension, one can easily see that the $4$-letter protocol
shows less oscillations in the low photon number regime (TMC and TWB)
than the $2$-letter one for a large range of loss value, at the price of
a slightly reduced capacity per letter.  On the other hand, no
appreciable improvement can be noticed for TTH states, thus confirming
that photon-number correlations carried by PNES are more effective for
quantum communication in lossy channel.  We conclude that, thanks to
their large stability, quaternary alphabets should be used in cases when
the mean photon number of the support cannot be precisely fixed. In a
$4$-letter protocol three thresholds should be used to extract the bit
values. The optimization shows that for TWB and TTH states the distances
between the three thresholds increase with the beam intensity whereas
for TMC states they are three consecutive numbers for any value of the
input energy. This behavior is due to the super-Poissonian
photon-number distributions of TWB and TTH.
\subsection{Asymmetric channels}
\begin{figure}[h!]
\begin{tabular}{lll}
\includegraphics[width=0.15\textwidth]{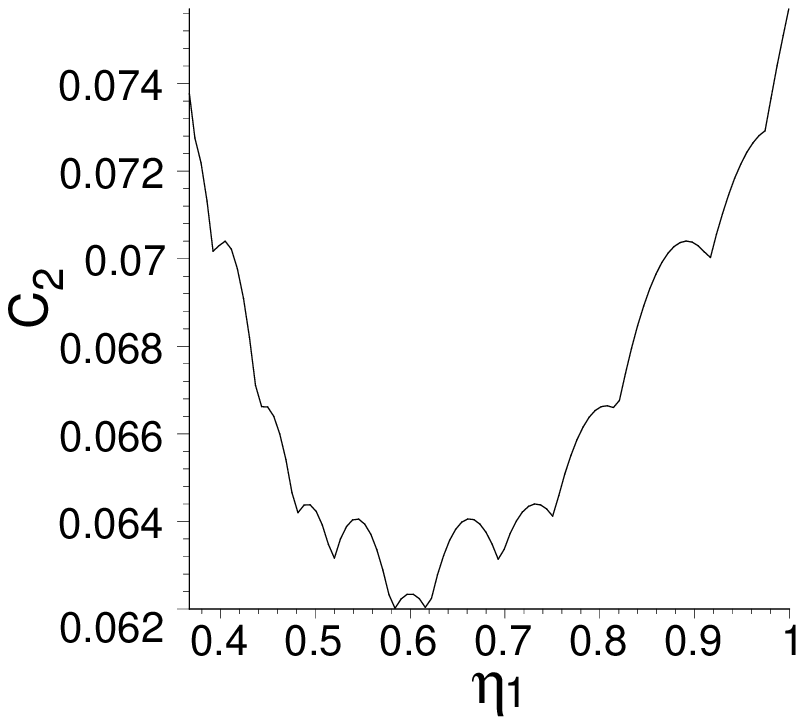}&
\includegraphics[width=0.15\textwidth]{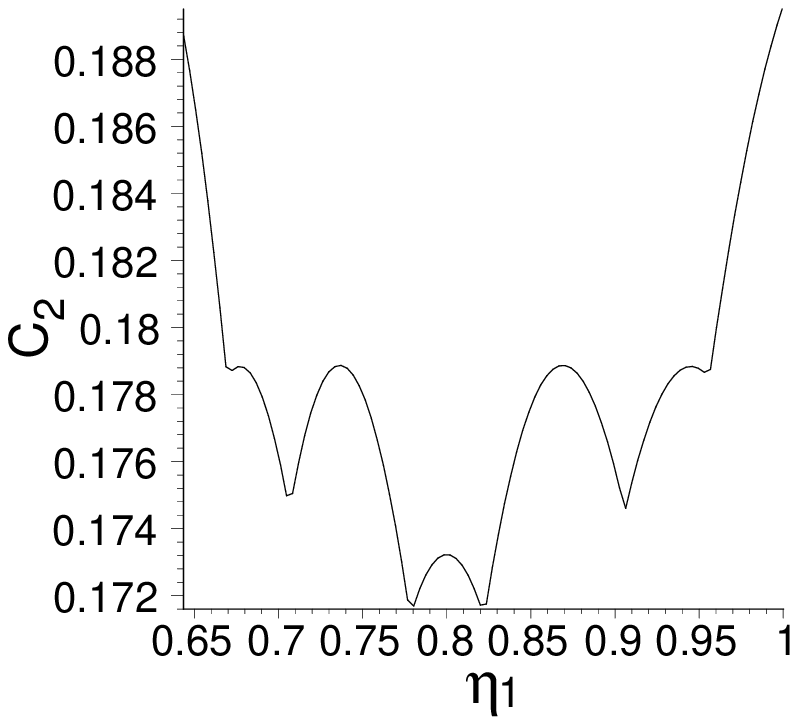}&
\includegraphics[width=0.15\textwidth]{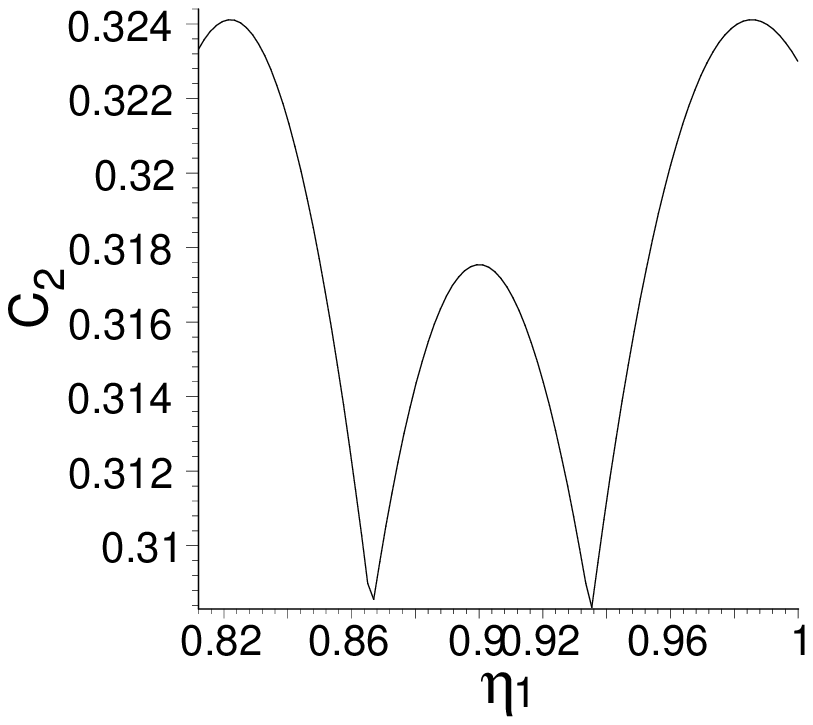}\\
\includegraphics[width=0.15\textwidth]{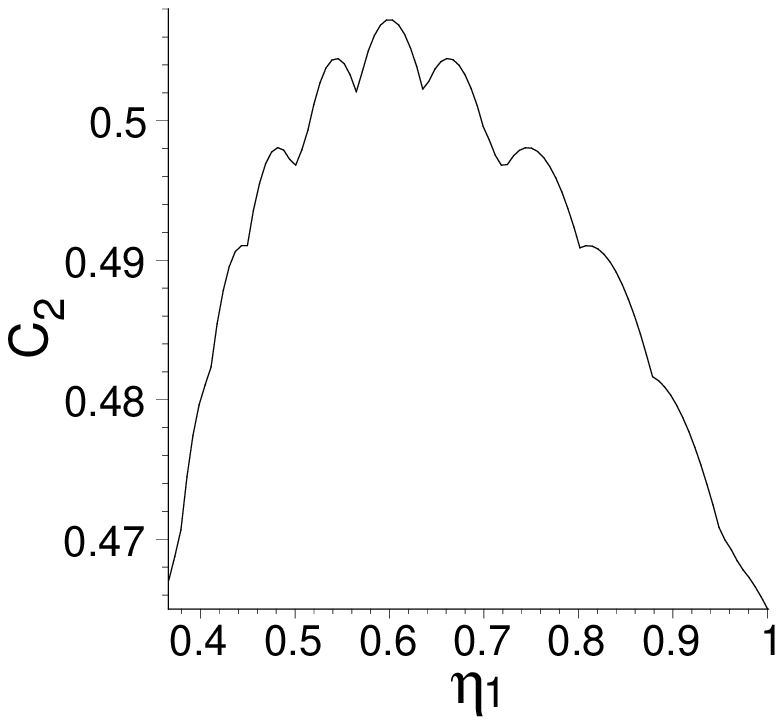}&
\includegraphics[width=0.15\textwidth]{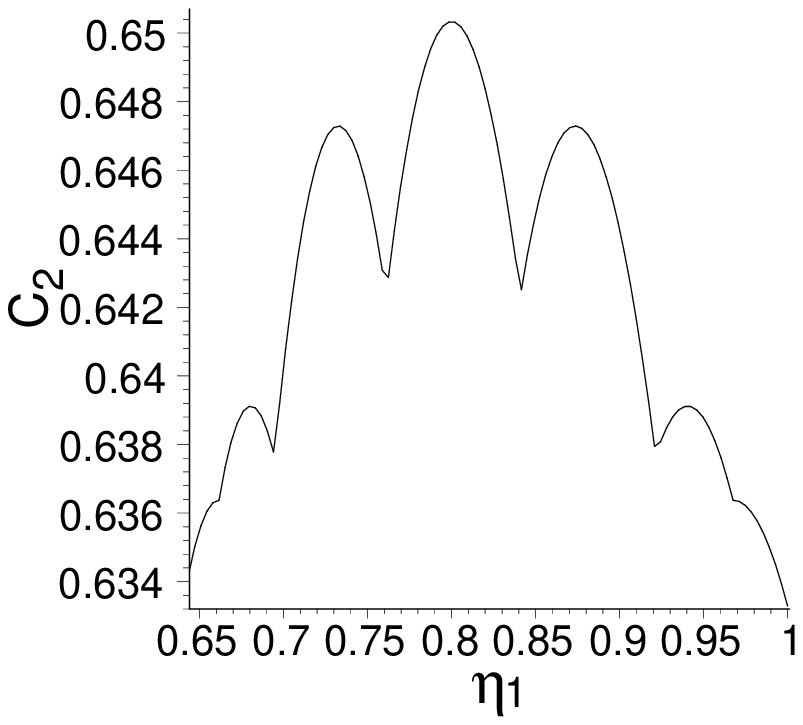}&
\includegraphics[width=0.15\textwidth]{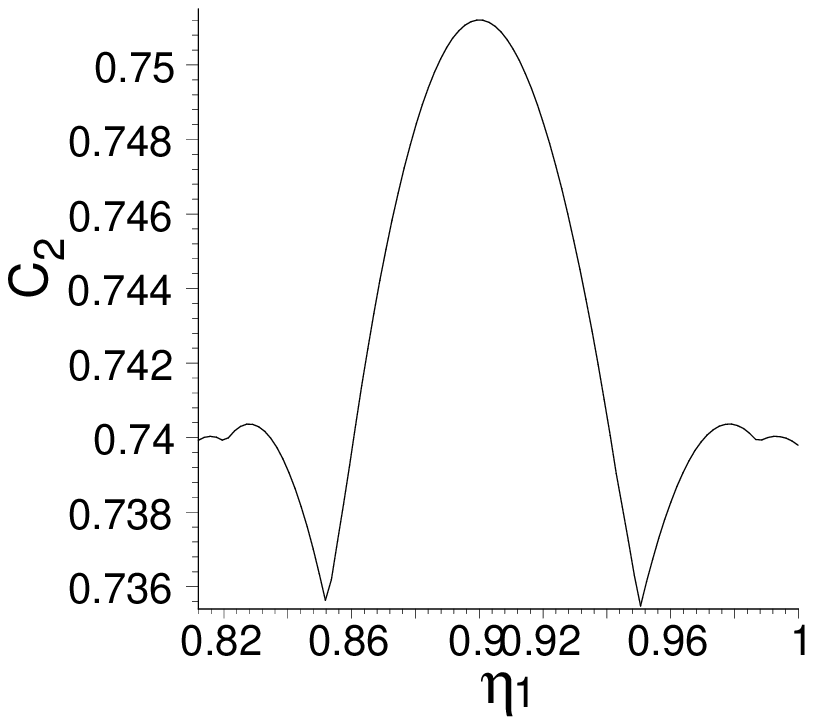}
\end{tabular}
\caption{Channel capacity for the TMC- (top) and TWB-based (bottom) 
2-letter CV coding in an asymmetric channel as a function of loss parameter 
in one of the channels. In all the plots the mode average photon number at the 
input is $\langle n \rangle = 10$. The overall channel loss is given, from 
left to right: $\eta=0.6$, $\eta=0.8$, $\eta=0.9$ 
\label{asym}}
\end{figure}

In this Section we analyze whether, and to which extent, different 
losses on the two beams affect the performances of the channel. In 
comparing symmetric to asymmetric channels we set the overall loss
$\eta=\sqrt{\eta_1\eta_2}$ and the beam energy and evaluate the bit 
threshold and the channel capacity by varying the asymmetry, {\em i.e} 
the loss of one channel, say $\eta_1$, in the range 
$\eta^2 \leq \eta_1\leq 1$. This scheme corresponds to set the 
overall distance between the two parties and move the source 
of PNES from one ($\eta_1=1$) to the other ($\eta_1=\eta^2$).
In Fig. \ref{asym} we show the channel capacities as a function of 
the single-channel loss $\eta_1$ for different values of the overall 
loss $\eta$ and a fixed value of the input beam energy. At first we notice 
that asymmetry is not dramatically affecting the performances of the 
channels, especially for the case of small overall loss ({\em i.e.} for
$\eta\rightarrow 1$). On the other hand, it is apparent from the plot that 
asymmetry acts in opposite way on the TMC- and TWB-based protocols. In 
fact, the channel capacity increases with asymmetry for TMC and decreases 
for TWB. This behavior depends on the different correlation properties 
of TMC and TWB. On the one hand, the correlation index of TMC remains 
unchanged  [compare to (\ref{gml})] in asymmetric channels whereas 
that of TWB decreases. On the other hand, asymmetry acts in opposite 
ways on the probability of coincidence counts in the two channels.
Upon expanding Eqs. (\ref{petaTMC}) and (\ref{petaTWB}) up to 
second order in the asymmetry we have 
\begin{align}
P_{\lambda,\eta_1,\eta_2} (n,n) & = P_{\lambda,\eta,\eta} (n,n)  
+ A_n(\eta,\lambda) \: \delta\eta^2   \\ 
P_{x,\eta_1,\eta_2} (n,n) & = P_{x,\eta,\eta} (n,n)  + 
B_n(\eta,x)\: \delta\eta^2  
\end{align}
where $\delta\eta = \eta_1 - \eta_2$, 
$A>0 \: \forall \eta,\lambda, n$ and $B<0 \: \forall \eta,x, n$.
Overall, placing the source of entanglement closer to one
of the two parties results in a slight increase of the capacity 
for TMC and a slight decrease for TWB.
\section{Discussion and conclusions}
\label{s:outro}
A question arises on whether security may be proved for PNES based
communication channels. In the ideal case of no loss, the
intercept-resend strategy has been considered assuming that Eve is able
to produce strongly  correlated beams source (optimally the TMC-source
\cite{tmc}) and it has been shown that the state-cloning attempts can be
revealed by checking the beam statistics, which is modified from
sub-Poissonian to super-Poissonian by any eavesdropping attempt. Since,
as we have proved in this paper, the statistical properties are not
changed by the propagation, TMC-based protocols are  secure also in the
presence of loss.  As concern TWB, security, remarkably security against
intercept-resend attacks, cannot be guaranteed through a check of the
beam statistics.  The TWB-based protocols require the use of additional
degrees of freedom, as for example binary randomization of polarization
\cite{twb} to guarantee security and to reveal eavesdropping actions.
This may also be useful for TMC-based protocols, in order to achieve
unconditional security.  Overall, there is a trade-off between the
quantity of information one is able to transmit at fixed energy and the
security of this transmission, with TMC offering more security at the
price of decreasing the channel capacity.
\par
In conclusion, we have analyzed lossy communication channels based on
photon number entangled states and realized upon choosing a shared set
of thresholds to convert the outcome of a joint photon number
measurement into a symbol from a binary or a quaternary alphabet.  We
have focused on channels build using two-mode coherently-correlated or
twin-beam states a support. The explicit optimization of the bit
discrimination thresholds have been performed and the corresponding
channel capacities have been compared to that of channels built using
classically correlated (separable) states.  We found that PNES are
useful to improve capacity in the presence of noise, and that TWB-based
channels may transmit a larger amount of information than TMC-based ones
at fixed energy and overall loss.  
\par
The evolution of the entangled support, either TMC or TWB, in lossy
channels have been analyzed in details, showing that the beam
statistics, either  sub-Poissonian for TMC or super-Poissonian for TWB,
is not altered during propagation. The preservation of sub-Poissonian
statistics indicates that TMC-based protocols are secure against
intercept-resend eavesdropping attacks, whereas TWB-based protocols
require the use of additional degrees of freedom, as for example binary
randomization of polarization.  
\par
We have analyzed the effects of asymmetric losses on the two beams,
showing that i) asymmetry of the channel does not dramatically affect
the performances and ii) placing the source of entanglement closer to
one of the two parties results in a slight increase of the capacity for
TMC-based protocols and a slight decrease for TWB-based ones.
\par
We conclude that photon-number entangled states, either Gaussian or non
Gaussian ones, are useful resources to implement effective
quantum-enhanced communication channels in the presence of loss.
\section*{Acknowledgments} This work has been supported by MIUR through the
project PRIN-2005024254-002. The work of VU has been supported by the Landau 
Network through the Cariplo Foundation fellowship program and by NATO through
grant CPB.NUKR.EV 982379. 

\end{document}